\documentstyle[twocolumn,psbox,myletter]{jpsj}

\def\runtitle{Spin Fluctuation-Induced Superconductivity in Organic Compounds}
\def\runauthor{Hisashi {\sc Kondo} and T\^{o}ru {\sc Moriya}}

\title
{
Spin Fluctuation-Induced Superconductivity in Organic Compounds
}

\author
{ 
Hisashi {\sc Kondo} and T\^{o}ru {\sc Moriya}
}

\inst
{
Department of Physics, 
Faculty of Science and Technology, Science University of Tokyo, Noda, 278-8510
}

\recdate
{
\today
}

\abst
{
    Spin fluctuation-induced superconductivity in two-dimensional organic 
compounds such as $\kappa$-$({\rm ET})_2X$ is investigated by using a simplified
dimer Hubbard model with right-angled isosceles triangular lattice
(transfer matrices $-{\mit \tau}$, $-{\mit \tau}^{\prime}$). The dynamical
susceptiblity and the self-energy are calculated self-consistently within the
fluctuation exchange approximation and the value for $T_{\rm c}$ as obtained by
solving the linearized Eliashberg-type equations is in good agreement with
experiment. The pairing symmetry is of ${\rm d}_{x^2-y^2}$ type. The calculated
$\left( U/{\mit \tau} \right)$-dependence of $T_{\rm c}$ compares qualitatively
well with the observed pressure dependence of $T_{\rm c}$. Varying the value for
${\mit \tau}^{\prime}/{\mit \tau}$ from 0 to 1  we interpolate
between the square lattice and the regular triangular lattice and find firstly
that values of $T_{\rm c}$ for $\kappa$-$({\rm ET})_2X$ and cuprates scale well
and secondly that $T_{\rm c}$ tends to decrease with increasing
${\mit \tau}^{\prime}/{\mit \tau}$ and no superconductivity is
found for ${\mit \tau}^{\prime}/{\mit \tau} = 1$, the regular
triangular lattice.
}

\kword
{
organic superconductor, spin fluctuation, Hubbard model, triangular lattice
}

\begin{document}
\sloppy
\maketitle


     Major topics in condensed matter physics in recent years may certainly
include the unconventional superconductivity observed in high-$T_{\rm c}$
cuprates, heavy electron systems and organic compounds. Among various possible
mechanisms proposed so far the spin fluctuation mechanism seems to have
possibilities of explaining all of these three.~\cite{rf:1} The superconductivity
in heavy electron systems is believed to be almost certainly due to the spin
fluctuation mechanism. Although the mechanism for cuprates is still controversial,
the spin fluctuation mechanism seems to be rather unique in its ability of
explaining not only the values of $T_{\rm c}$ and the pairing symmetry but
also various
anomalous physical properties in the normal state as well as in the
superconducting state in a systematic fashion at least in the optimal and
overdoped concentration regimes;~\cite{rf:2} the underdoped regime still remains
to be clarified.
     
     As for the two-dimensional (2D) organic superconductors the spin fluctuation
mechanism seems to be the only available mechanism provided the superconducting
order parameter is anisotropic, say of d-wave, as was indicated by recent
investigations.~\cite{rf:3, rf:4, rf:5} Major differences of this problem from
that of high-$T_{\rm c}$ cuprates are that the superconductivity occurs without
doping and in many cases in the metallic side of a metal-insulator Mott
transition and thus the system should be in the intermediate correlation regime.
For example, the $t$-$J$ model should safely be excluded. 
     
     We wish to discuss the spin fluctuation mechanism of superconductivity in
2D organic compounds, keeping
$\kappa$-$({\rm ET})_2X$ $[{\rm ET} = {\rm BEDT}$-${\rm TTF}, 
      X = {\rm Cu} \{ {\rm N(CN)_2} \} X^{\prime}, X^{\prime} = {\rm Cl},
      {\rm Br}]$
in mind.~\cite{rf:6} According to experiment $\kappa$-$({\rm ET})_2X$
$[X^{\prime} = {\rm Cl}]$ is an antiferromagnetic insulator and under increasing
pressure it undergoes a insulator to superconductor transition.~\cite{rf:7}
Each layer of
molecules in these compounds may be regarded as consisting of dimers each of
which has one hole in the antibonding dimer orbital of the highest occupied
molecular orbitals (HOMO).~\cite{rf:8, rf:9} There are transfer matrices
$- {\mit \tau} \left( {\mit \tau} > 0 \right)$ between
dimers and the additional consideration of intra-dimer electron interaction $U$
naturally leads to the Hubbard model. For $U \gg {\mit \tau}$ we have
antiferromagnetic ground state and with decreasing $U/{\mit \tau}$ we encounter
an insulator to metal Mott transition at a certain value of $U/{\mit \tau}$.

     We now wish to study possible superconductivity on the metallic side of
this transition. For this purpose we make use of a half-filled single band
Hubbard model consisting of antibonding dimer orbitals with the inter-dimer
transfers and the intra-dimer electron interaction. The spin fluctuations are
treated within the renormalized random phase approximation (RRPA) or the
fluctuation exchange (FLEX) approximation.~\cite{rf:10} Although the formalism
of the
self-consistent renormalization (SCR) theory takes into account the vertex
corrections,~\cite{rf:11} associated numerical jobs seem to be too heavy to
carry them
through. The results of FLEX for cuprates so far seem to be fairly
successful and there are a few arguments on vertex corrections in support of
the FLEX approach.~\cite{rf:2, rf:12, rf:13, rf:14, rf:15, rf:21, rf:16} 

     Thus we see that the problem itself is reduced to the one quite similar to
the Hubbard model description of cuprates. Only difference is the lattice
structure and the transfer matrices. As a matter of fact we may simplify, to a
good approximation, the model for ${\mit \kappa}$-$({\rm ET})_2 X$ to a square
lattice with the nearest neighbor transfers ${\mit \tau}$ and one of the cross
diagonal second neighbor transfers ${\mit \tau}^{\prime}$ (say up right corner
to down left), thus making a right-angled isosceles triangular lattice with
transfer matrices $-{\mit \tau}$ for the two sides and $-{\mit \tau}^{\prime}$ for
the base, Fig. 1(a).~\cite{rf:17} The value for
\begin{figure}[t]
  \begin{center}
    \psbox[size=0.55#1]{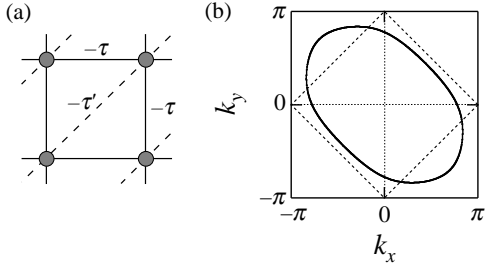}
  \end{center}
  \caption{
  (a) The model unit cell and the transfer integrals.
  (b) Unperturbed Fermi surface for ${\mit \tau}^{\prime} / {\mit \tau} = 0.8$.
  Dashed lines show the antiferromagnetic zone boundary.
  }
\end{figure}
${\mit \tau}^{\prime}/{\mit \tau}$ as
estimated from the presently accepted values for the transfer integrals is
about 0.8 (close to 1) for ${\mit \kappa}$-$({\rm ET})_2 X$.~\cite{rf:8, rf:9}
This reminds us of the magnetic frustration problem in an equilateral triangular
lattice. So we
study not only the case of  ${\mit \kappa}$-$({\rm ET})_2 X$ but also the
problem for varying values of ${\mit \tau}^{\prime}/{\mit \tau}$, including
the regular triangle: ${\mit \tau}^{\prime}/{\mit \tau} = 1$. 

     The model Hamiltonian is given by
\begin{eqnarray}
     H &=& - \sum_{\mit \sigma} \sum_{\langle i,j \rangle} 
          {\mit \tau} a^{\dagger}_{i {\mit \sigma}} a_{j {\mit \sigma}}
         - \sum_{\mit \sigma} \sum_{\left( i,j \right)} 
          {\mit \tau}^{\prime} a^{\dagger}_{i {\mit \sigma}} a_{j {\mit \sigma}}
\nonumber  \\    && ~~~~~~~~~~~~~~~~~~~~~~~~ 
       + U \sum_{j} n_{j \uparrow} n_{j \downarrow}
\label{eq:1}
\\
      &=& - \sum_{\mit \sigma} \sum_{\mibs k} 
          {\mit \epsilon}_{\mibs k}
                 a^{\dagger}_{{\mibs k} {\mit \sigma}} a_{{\mibs k} {\mit \sigma}}
       + U \sum_{j} n_{j \uparrow} n_{j \downarrow}, 
\label{eq:2}
\end{eqnarray}
with
\begin{eqnarray}
    {\mit \epsilon}_{\mibs k} &=& 
       - 2 {\mit \tau} \left( \cos k_{x} a + \cos k_{y} a \right) 
\nonumber  \\    && ~~~~~~~~~~~~~~~~~~  
       - 2 {\mit \tau}^{\prime} \cos \left( k_{x} a + k_{y} a \right) ,
\label{eq:3}
\end{eqnarray}
where $\langle i,j \rangle$ and $\left( i,j \right)$ indicate the nearest
neighbor pairs and the second neighbor side diagonal pairs, respectively, as
shown in Fig. 1(a) and $a$ is the lattice constant.
     
     The equations for the normal and anomalous Green's functions are given by
\begin{eqnarray}
     G \left( {\mib k} , {\rm i} {\mit \omega}_n \right) &=&
       G^{\left( 0 \right)} \left( {\mib k} , {\rm i} {\mit \omega}_n \right)
\nonumber  \\    && +
       G^{\left( 0 \right)} \left( {\mib k} , {\rm i} {\mit \omega}_n \right)
       \left[ {\mit \Sigma}^{\left( 1 \right)} 
                       \left( {\mib k} , {\rm i} {\mit \omega}_n \right)
       G \left( {\mib k} , {\rm i} {\mit \omega}_n \right)  \right.
\nonumber  \\    &&  \left.
     -  {\mit \Sigma}^{\left( 2 \right)} 
                       \left( {\mib k} , {\rm i} {\mit \omega}_n \right)
       F^{*} \left( {\mib k} , {\rm i} {\mit \omega}_n \right) \right], 
\label{eq:4}
\\
     F^{*} \left( {\mib k} , {\rm i} {\mit \omega}_n \right) &=&
       G^{\left( 0 \right)} \left( -{\mib k} , -{\rm i} {\mit \omega}_n \right)
\nonumber  \\    && \times
       \left[ {\mit \Sigma}^{\left( 1 \right)} 
                       \left( - {\mib k} , - {\rm i} {\mit \omega}_n \right)
       F^{*} \left( {\mib k} , {\rm i} {\mit \omega}_n \right) \right.
\nonumber  \\    && \left.
      + {\mit \Sigma}^{\left( 2 \right)} 
                       \left( - {\mib k} , - {\rm i} {\mit \omega}_n \right)
       G \left( {\mib k} , {\rm i} {\mit \omega}_n \right) \right],
\label{eq:5}
\end{eqnarray}
where $G^{\left( 0 \right)} \left( {\mib k} , {\rm i} {\mit \omega}_n \right)$
is the Green's function for a non-interacting system and the self-energies due
to the spin and charge fluctuations are given as follows:
\begin{eqnarray}
     {\mit \Sigma}^{\left( 1 \right)}
                   \left( {\mib k} , {\rm i} {\mit \omega}_n \right) &=&
       \frac{T}{N_0} \sum_{{\mibs q} , m}
         V^{\left( 1 \right)} \left( {\mib q} , {\rm i} {\mit \Omega}_m \right) 
\nonumber  \\ && ~~~~~~~~~~ \times
         G \left( {\mib k}-{\mib q} , 
                     {\rm i} {\mit \omega}_n - {\rm i} {\mit \Omega}_m \right),
\label{eq:6}
\\
     {\mit \Sigma}^{\left( 2 \right)}
                   \left( {\mib k} , {\rm i} {\mit \omega}_n \right) &=&
       - \frac{T}{N_0} \sum_{{\mibs q} , m}
         V^{\left( 2 \right)} \left( {\mib q} , {\rm i} {\mit \Omega}_m \right) 
\nonumber  \\ && ~~~~~~~~~~ \times
         F^{\left( 2 \right)} \left( {\mib k}-{\mib q} , 
                     {\rm i} {\mit \omega}_n - {\rm i} {\mit \Omega}_m \right),
\label{eq:7}
\end{eqnarray}
with 
\begin{eqnarray}
     V^{\left( 1 \right)}
                   \left( {\mib q} , {\rm i} {\mit \Omega}_m \right) &=&
       U + U^2 \left[ 
          \frac{3}{2}
           {\mit \chi}_{\rm s} \left( {\mib q} , {\rm i} {\mit \Omega}_m \right)
         +\frac{1}{2}
           {\mit \chi}_{\rm c} \left( {\mib q} , {\rm i} {\mit \Omega}_m \right)
                      \right.
\nonumber  \\    &&        \left.
         -\frac{1}{2} \left\{ 
           {\overline {\mit \chi}}_{\rm s}
                     \left( {\mib q} , {\rm i} {\mit \Omega}_m \right)
          +{\overline {\mit \chi}}_{\rm c}
                     \left( {\mib q} , {\rm i} {\mit \Omega}_m \right)
             \right\} \right],
\label{eq:8}
\\
     V^{\left( 2 \right)}
                   \left( {\mib q} , {\rm i} {\mit \Omega}_m \right) &=&
       U + U^2 \left[ 
          \frac{3}{2}
           {\mit \chi}_{\rm s} \left( {\mib q} , {\rm i} {\mit \Omega}_m \right)
         -\frac{1}{2}
           {\mit \chi}_{\rm c} \left( {\mib q} , {\rm i} {\mit \Omega}_m \right)
                      \right.
\nonumber  \\    &&        \left.
         -\frac{1}{2} \left\{ 
           {\overline {\mit \chi}}_{\rm s}
                     \left( {\mib q} , {\rm i} {\mit \Omega}_m \right)
          -{\overline {\mit \chi}}_{\rm c}
                     \left( {\mib q} , {\rm i} {\mit \Omega}_m \right)
             \right\} \right],
\label{eq:9}
\end{eqnarray}
and
\begin{eqnarray}
     {\mit \chi}_{\rm s} \left( {\mib q} , {\rm i} {\mit \Omega}_m \right) &=&
       \frac{{\overline {\mit \chi}}_{\rm s}
                     \left( {\mib q} , {\rm i} {\mit \Omega}_m \right)}
            {1 - U {\overline {\mit \chi}}_{\rm s}
                     \left( {\mib q} , {\rm i} {\mit \Omega}_m \right)}, 
\nonumber  \\
     {\mit \chi}_{\rm c} \left( {\mib q} , {\rm i} {\mit \Omega}_m \right) &=&
       \frac{{\overline {\mit \chi}}_{\rm c}
                     \left( {\mib q} , {\rm i} {\mit \Omega}_m \right)}
            {1 + U {\overline {\mit \chi}}_{\rm c}
                     \left( {\mib q} , {\rm i} {\mit \Omega}_m \right)},
\label{eq:10}
\end{eqnarray}
\begin{eqnarray}
    {\overline {\mit \chi}}_{\rm s} 
                 \left( {\mib q} , {\rm i} {\mit \Omega}_m \right) &=&
      - \frac{T}{N_0} \sum_{{\mibs k} , n} \left[
         G \left( {\mib k}+{\mib q} , 
                     {\rm i} {\mit \omega}_n + {\rm i} {\mit \Omega}_m \right) 
         G \left( {\mib k} , {\rm i} {\mit \omega}_n \right) \right.
\nonumber  \\
      && \left.
       + F \left( {\mib k}+{\mib q} , 
                     {\rm i} {\mit \omega}_n + {\rm i} {\mit \Omega}_m \right) 
         F \left( {\mib k} , {\rm i} {\mit \omega}_n \right) \right],
\label{eq:11}
\\
    {\overline {\mit \chi}}_{\rm c} 
                 \left( {\mib q} , {\rm i} {\mit \Omega}_m \right) &=&
      - \frac{T}{N_0} \sum_{{\mibs k} , n} \left[
         G \left( {\mib k}+{\mib q} , 
                     {\rm i} {\mit \omega}_n + {\rm i} {\mit \Omega}_m \right) 
         G \left( {\mib k} , {\rm i} {\mit \omega}_n \right) \right.
\nonumber  \\
      && \left.
       - F \left( {\mib k}+{\mib q} , 
                     {\rm i} {\mit \omega}_n + {\rm i} {\mit \Omega}_m \right) 
         F \left( {\mib k} , {\rm i} {\mit \omega}_n \right) \right],
\label{eq:12}
\end{eqnarray}
where ${\mit \omega}_n = \left( 2n+1 \right) {\mit \pi} T$ and 
${\mit \Omega}_m = 2m {\mit \pi} T$ are the Fermi and the Bose Matsubara
frequencies, respectively and $N_0$ is the number of dimers in the crystal. 
We study have the possibility of singlet pairing.

     Confining ourselves here to the transition temperature $T_{\rm c}$ and the
normal state properties we may linearize the equations (\ref{eq:4}, \ref{eq:5})
and (\ref{eq:11}, \ref{eq:12}) with
respect to the anomalous self-energy
${\mit \Sigma}^{\left( 2 \right)}
                   \left( {\mib k} , {\rm i} {\mit \omega}_n \right)$ or 
$F^{\left( 2 \right)} \left( {\mib k} , {\rm i} {\mit \omega}_n \right)$. 
$T_{\rm c}$ can be calculated as the highest temperature where the following
equation for the normalized anomalous self-energy
$f \left( {\mib k} , {\rm i} {\mit \omega}_n \right) = 
       {\mit \Sigma}^{\left( 2 \right)}
                   \left( {\mib k} , {\rm i} {\mit \omega}_n \right)
       \left| G \left( {\mib k} , {\rm i} {\mit \omega}_n \right) \right|$
has a non-trivial solution:
\begin{eqnarray}
  &&   f \left( {\mib k} , {\rm i} {\mit \omega}_n \right) 
\nonumber  \\
  && ~~ = -\frac{ T U }{N_0}
       \sum_{{\mibs p} , m}
       \left| G \left( {\mib k} , {\rm i} {\mit \omega}_n \right) \right|
       \left[ 1+\frac{3}{2} U
           {\mit \chi}_{\rm s} \left( {\mib k} - {\mib p} , 
                   {\rm i} {\mit \omega}_n - {\rm i} {\mit \omega}_m \right) 
                                     \right.
\nonumber  \\ && ~~~~~~~~~~~~~ \left.
         -\frac{1}{2} U
           {\mit \chi}_{\rm c} \left( {\mib k} - {\mib p} , 
                {\rm i} {\mit \omega}_n - {\rm i} {\mit \omega}_m \right) \right]
\nonumber  \\ && ~~~~~~~~~~~~~~~~~~ \times
       \left| G \left( {\mib p} , {\rm i} {\mit \omega}_m \right) \right|
       f \left( {\mib p} , {\rm i} {\mit \omega}_m \right).
\label{eq:13}
\end{eqnarray}
The symmetry of the order parameter is given by that of
$f \left( {\mib k} , {\rm i} {\mit \omega}_n \right)$. Since the kernel of
eq. (\ref{eq:13}) has a full symmetry of the model
Hamiltonian we may classify the order
parameter according to the irreducible representations of the symmetry group.
There are four one-dimensional irreducible representations ${\rm A}_1$,
${\rm A}_2$, ${\rm B}_1$ and ${\rm B}_2$.  We take advantage of using this
symmetry property in the calculation.~\cite{rf:18} For numerical calculations 
we take $64 \times 64$ ${\mib k}$-points and 512 and 2048 Matsubara frequencies
for susceptibility and self-energy, respectively.

     We first show the results of calculation for
${\mit \kappa}$-$({\rm ET})_2 X$ 
$\left( {\mit \tau}^{\prime}/{\mit \tau} = 0.8 \right)$.
The unperturbed Fermi surface is shown in Fig. 1(b).
\begin{figure}[t]
  \begin{center}
    \psbox[size=0.70#1]{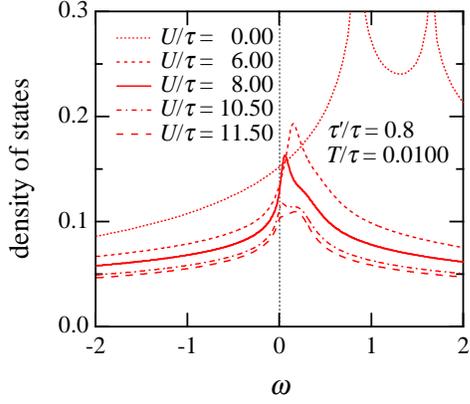}
  \end{center}
  \caption{
  Density of states for various values of $U / {\mit \tau}$.
  }
\end{figure}
\begin{figure}[t]
  \begin{center}
    \psbox[size=0.65#1]{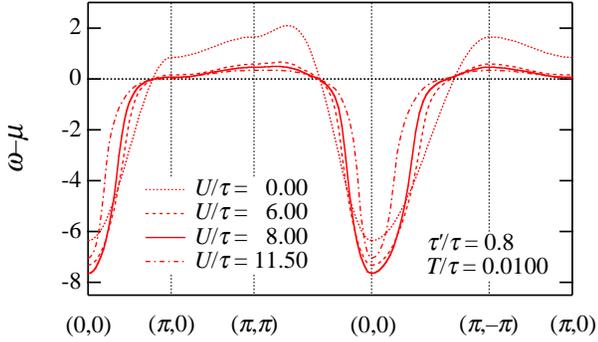}
  \end{center}
  \caption{
  Quasi-particle dispersion.
  }
\end{figure}
\begin{figure}[t]
 \begin{center}
    \psbox[size=0.65#1]{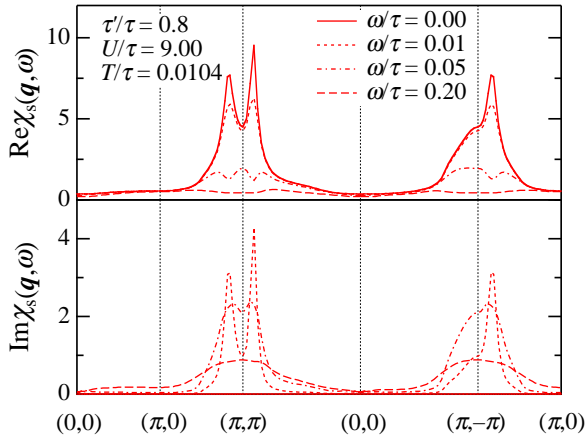}
  \end{center}
  \caption{
  Dynamical susceptibility.
  }
\end{figure}
The one particle density of states and the
quasi-particle dispersions are shown in Figs. 2 and 3, respectively, together
with the corresponding values for $U = 0$. The dynamical susceptibilities are
shown in Fig. 4 for various frequencies. We see that incommensurate peaks
around $({\mit \pi},{\mit \pi})$ and an asymmetric peak around
$({\mit \pi},-{\mit \pi})$ are strongly enhanced. 

     The superconducting transition temperature $T_{\rm c}$ is given as
$T_{\rm c} = 0.0105 {\mit \tau}$ for $U = 9 {\mit \tau}$ and the order parameter
has the ${\rm A}_2$ symmetry $\left( x^2 - y^2 \right)$. Estimating 
${\mit \tau} \sim 0.07 {\rm eV}$, we get $T_{\rm c} \sim 9 {\rm K}$ in fair
agreement with the observed value $T_{\rm c} \sim 10 {\rm K}$. Fig. 5 shows the
\begin{figure}[t]
  \begin{center}
    \psbox[size=0.75#1]{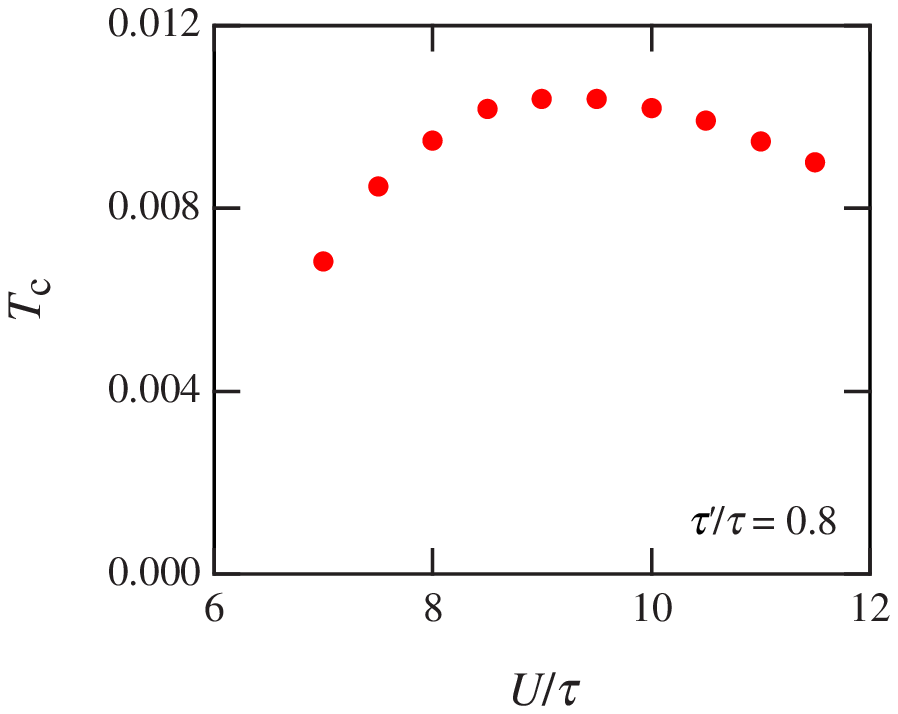}
  \end{center}
  \caption{
  $T_{\rm c}$ vs. $U / {\mit \tau}$.
  }
\end{figure}
plot of $T_{\rm c}$ vs. $U / {\mit \tau}$. $T_{\rm c}$ shows a weak maximum at 
$U / {\mit \tau} \sim 9$ and then decreases with
decreasing $U / {\mit \tau}$. Since $U / {\mit \tau}$ should decrease with
increasing pressure, this tendency is consistent with the observed pressure
dependence of $T_{\rm c}$ if we assume a proper critical value
$\left( U / {\mit \tau} \right)_{\rm c}$, say around 10,
for the insulator-superconductor
transition.~\cite{rf:7, rf:19} According to the mean field calculation,
which normally
underestimate the critical point for transition, the first order transition
between the insulator and metallic phases takes place at
$U / {\mit \tau} = 4.2$.~\cite{rf:9} It still remains to calculate the critical 
value $\left( U / {\mit \tau} \right)_{\rm c}$ by using the same FLEX
approximation for both antiferromagnetic and superconducting states at $T = 0$.

     We also note that the calculated values of $T_{\rm c}$ for
${\mit \kappa}$-$({\rm ET})_2 X$ and cuprates properly scale. A crude model
for some of the cuprates, say LSCO, may correspond to the case of 
${\mit \tau}^{\prime}/{\mit \tau} \approx 0$ with less than
half-filled carriers. According
to the previous calculations we have 
$T_{\rm c}/{\mit \tau}= 0.0273 \left( 0.021 \right)$ for 
${\mit \tau}^{\prime}/{\mit \tau} = 0$ (0.15 for cross diagonal transfers), 
$U/{\mit \tau} = 6 \left( 4 \right)$ and the carrier number
$n = 0.875$.~\cite{rf:12, rf:13}
Since the band width of cuprates is considered to be roughly $2 \sim 4$ times
larger than that in ${\mit \kappa}$-$({\rm ET})_2 X$, the calculated relative
values of $T_{\rm c}$ compare rather well with experiment. 

     Next we discuss the
$\left( {\mit \tau}^{\prime}/{\mit \tau} \right)$-dependence
of $T_{\rm c}$ for the half-filled case. It is interesting to find that
eq. (\ref{eq:13}) has no solution for ${\mit \tau}^{\prime}/{\mit \tau} = 1$
(the regular
triangular lattice) in any reasonable range of the value for $U/{\mit \tau}$,
or $U/{\mit \tau} < 16$, the highest value studied. For $U/{\mit \tau} \gg 1$ and 
${\mit \tau}^{\prime}/{\mit \tau} = 1$ we have an antiferromagnetic Heisenberg
model with a regular triangular lattice, a famous frustrated system.~\cite{rf:20}
Around the Mott transition under pressure the local moments disappear and the
metallic phase seems to be characterized by a wave vector dependent magnetic
susceptibility with three broad peaks as is shown in Fig. 6.  This situation
\begin{figure}[t]
  \begin{center}
    \psbox[size=0.65#1]{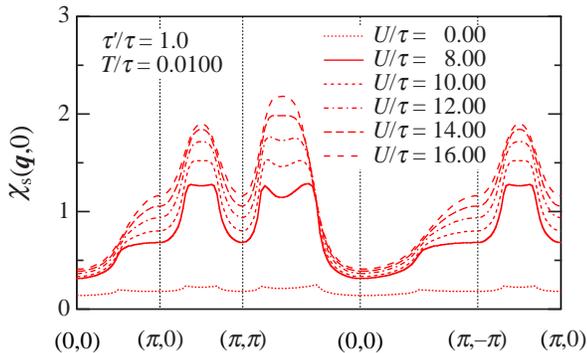}
  \end{center}
  \caption{
  Wave vector dependence of the susceptibility for
  ${\mit \tau}^{\prime} / {\mit \tau} = 1.0$, the regular triangular lattice, 
  for various values of $U / {\mit \tau}$.
  }
\end{figure}
does not seem favorable for the spin fluctuation-induced superconductivity
as may be seen from a weak coupling argument.~\cite{rf:22}

     In the case of ${\mit \tau}^{\prime}/{\mit \tau} = 0.4$,
the antiferromagnetic
peak of the dynamical susceptibility is more significant than for 
${\mit \tau}^{\prime}/{\mit \tau} = 0.8$ and the superconductivity appears for
smaller $U$ and the value of $T_{\rm c}$ is higher as may naturally be expected.
For example, we have $T_{\rm c}/{\mit \tau}=0.014$ and $0.016$
for $U/{\mit \tau}=3.3$ and $3.7$, respectively.
For a half-filled band we may conjecture as follows: Considering a phase diagram 
in $U/{\mit \tau}$ vs. ${\mit \tau}^{\prime}/{\mit \tau}$ plane, the critical 
boundary for antiferromagnetism $\left( U/{\mit \tau} \right)_{\rm AF}$ is lower
than that for superconductivity $\left( U/{\mit \tau} \right)_{\rm SC}$ for small
values of ${\mit \tau}^{\prime}/{\mit \tau}$ where the nesting condition is well 
satisfied. With increasing ${\mit \tau}^{\prime}/{\mit \tau}$ the frustration 
increases and its destructive influence is stronger for antiferromagnetism than
for superconductivity and thus $\left( U/{\mit \tau} \right)_{\rm SC}$
becomes lower than 
$\left( U/{\mit \tau} \right)_{\rm AF}$ beyond a certain value of 
${\mit \tau}^{\prime}/{\mit \tau}$. For a definite conclusion it is necessary to
compare the free energies of both of these states.

     As for the normal state properties the uniform susceptibility is almost
constant and has weak tendency of decreasing with decreasing temperature.
The calculated values for $\sum_{\mibs q} {\rm Im} {\mit \chi}
                  \left( {\mib q} ,{\mit \omega} \right) / {\mit \omega}$,
the quantity proportional to $1 / T_1 T$, $T_1$ being the nuclear spin-lattice
relaxation rate, has a Curie-Weiss like temperature dependence down to
$T_{\rm c}$. On the other hand the experimental results on 
${\mit \kappa}$-$({\rm ET})_2X$ $[X^{\prime} = {\rm Br}]$ show a substantial
decrease of the susceptibility as $T_{\rm c}$ is approached and the peak
structure of $1 / T_1 T$ much above $T_{\rm c}$.~\cite{rf:3, rf:4, rf:5}
These behaviors remain to be explained. 

     In summary we have studied the spin fluctuation mechanism of 
superconductivity in ${\mit \kappa}$-$({\rm ET})_2X$ by using a FLEX
approximation for a dimer Hubbard model with the effective transfer integrals 
$-{\mit \tau}$, $-{\mit \tau}^{\prime}$ deduced from the presently accepted
values for the transfer integrals and varying values of $U$.
We found superconductivity of ${\rm d}_{x^2-y^2}$ types. The value for $T_{\rm c}$
and its $U/{\mit \tau}$ (pressure) dependence compare well with experiments.
We point out that if we vary the value of ${\mit \tau}^{\prime}/{\mit \tau}$
in the present model for ${\mit \kappa}$-$({\rm ET})_2X$ between 0 and 1 the
square lattice and the regular triangular lattice Hubbard models are 
interpolated. We find that the values of $T_{\rm c}$ for 
${\mit \kappa}$-$({\rm ET})_2X$ and for cuprates scale fairly well and there is
no superconductivity 
of singlet pairing
for the triangular lattice within a reasonable range of
the value for $U$. However, the anomalous normal state properties in the uniform
susceptibility and the nuclear spin-lattice relaxation rate, bearing
resemblance to the pseudo-spingap phenomena in cuprates, remain to be explained.
Also, the transition between the antiferromagnetic insulator and superconducting
phases is still to be investigated.

    We would like to thank Prof. K. Kanoda, Dr. S. Nakamura and in particular 
Dr. T. Takimoto for helpful discussions.


\begin{thebibliography}{99}
\bibitem{rf:17}
    Strictly speaking, two dimers make a body-centered unit cell and the
    transfer integrals in the vertical and horizontal directions in Fig. 1 are
    different by 1.5 percent. We neglect this difference for substantial
    simplification of the model.
\end{thebibliography}
\end{document}